\documentclass[a4paper,fleqn,usenatbib]{mnras}

\usepackage[T1]{fontenc}
\usepackage{ae,aecompl}

\usepackage{graphicx}	
\usepackage{amsmath}	
\usepackage{amssymb}	

\title[LMC globular cluster kinematics]{Two kinematically distinct old globular cluster
populations in the Large Magellanic Cloud}

\author[A.E. Piatti, et al.]{
Andr\'es E. Piatti$^{1,2}$\thanks{E-mail: andres@oac.unc.edu.ar}, Emilio J. Alfaro$^3$, 
Tristan Cantat-Gaudin$^4$\\
$^{1}$Consejo Nacional de Investigaciones Cient\'{\i}ficas y T\'ecnicas, Godoy Cruz 2290, C1425FQB, 
Buenos Aires, Argentina\\
$^{2}$Observatorio Astron\'omico de C\'ordoba, Laprida 854, 5000, 
C\'ordoba, Argentina\\
$^3$Instituto de Astrof\'{\i}sica de Andaluc\'{\i}a, (CSIC), Glorieta de la Astronom\'{\i}a, S/N, Granada, 18008, Spain\\
$^4$Institut de Ci\`encies del Cosmos, Universitat de Barcelona (IEEC-UB), Mart\'{\i} i Franqu\`es 1, E-08028 Barcelona, Spain
}

\date{Accepted XXX. Received YYY; in original form ZZZ}

\pubyear{2018}

\begin{document}
\label{firstpage}
\pagerange{\pageref{firstpage}--\pageref{lastpage}}
\maketitle

\begin{abstract}
We report results of proper motions of 15 known Large Magellanic Cloud (LMC) old
globular clusters (GCs) derived from the {\it Gaia} DR2 data sets. When these mean proper motions are
gathered with existent radial velocity measurements to compose the GCs' velocity vectors,
we found that the projection of the velocity vectors onto the LMC plane and those
perpendicular to it tell us about two distinct kinematic GC populations. Such a distinction
becomes clear if the GCs are split at a perpendicular velocity of 10 km/s (absolute value). 
The two different kinematic groups also exhibit different spatial distributions. Those with smaller vertical velocities are part of the LMC disc, while those
with larger values are closely distributed like a spherical component. Since GCs in both
kinematic-structural components share similar ages and metallicities, we speculate with
the possibility that their origins could have occurred through a fast collapse that formed 
halo and disc concurrently. 
\end{abstract} 




\section{Introduction}

The orbital motions of the old globular clusters (GCs) in the Large Magellanic Cloud
(LMC) have been described from radial velocity (RV) measurements by a disc-like 
rotation with no GCs appearing to have halo kinematics \citep{s92,getal06,shetal10}, 
so that it is expected that they do not cross the LMC disc. However, a closer
look to the RV versus position angle (PA) diagram, tells us that although a general trend 
following the LMC disc rotation is visible, there
are also some GCs that clearly depart from that relationship. This behaviour can be seen
in the top-left panel of Fig.~\ref{fig:fig1}, built from RVs and PAs taken from
\citet{piattietal2018}. The curve for a LMC disc, rotating according to the solution found 
from  $HST$ proper motion measurements in 22 fields \citep{vdmk14}, is also depicted with 
a solid line. The dotted lines represent the curves considering the quoted errors in all the 
parameters involved
(e.g., inclination of the disc, PA of the line-of-nodes, LMC dynamical centre, disc
rotation velocity, etc) and the velocity dispersion, added in quadrature. 

This observational evidence, together with some speculations about their origins 
\citep[e.g.][]{brocatoetal1996,pg13}, their different kinematics with respect to the
LMC halo field stars \citep[][and references therein]{bekki2007}, the effects on their
motions because of the interaction of the LMC with the Milky Way \citep{bekki2011,carpinteroetal2013}, 
among others,
points to the need of deriving their space velocities, so that a three-dimensional
picture of their movements can be analysed. In order to obtain them, accurate proper motions
are necessary. With the second data release (DR2) of {\it Gaia} mission \citep{gaiaetal2016,gaiaetal2018b}, 
this challenging goal has became to be possible for the first time. Precisely, 
we derive here such LMC GCs' proper motions and from information gathered 
from the literature  constructed their space velocity vectors. 
By analysing their rotational and vertical velocities, along
with their relationships with the GCs' positions in the galaxy, their ages and metallicities, 
we found that two groups of GCs is possible to  be differentiated from a kinematics point of view.

The Letter is organised as follows: in Section 2 we describe the selection of the {\it Gaia} DR2
extracted data and the estimation of the GCs' mean proper motions. In Section 3 we perform a
rigorous transformation of RVs and proper motions in terms of the rotation in and perpendicular 
to the LMC plane and discuss the results to the light of possible GC formation scenarious.

\section{LMC GCs' proper motions}

From the {\it Gaia} archive\footnote{http://gea.esac.esa.int/archive/}, we extracted
parallaxes ($\varpi$) and proper motions in Right Ascension (pmra) and Declination (pmdec) 
for stars located within 10 arcmin from the centres of 15 known LMC GCs (see Table~\ref{tab:table1}).
We limited our sample
to stars with proper motion errors $\le$ 0.5 mas/yr, which correspond to 
$G$ $\la$ 19.0 mag \citep{gaiaetal2018a}.  Excess noise (\texttt{epsi}) and 
significance of excess of noise (\texttt{sepsi}) in Gaia DR2 \citep{lindegrenetal2018} 
were used to 
prune the data. D(=\texttt{sepsi}) $<$ 2 and \texttt{epsi} $<$ 1 define a good balance between data quality and number of retained objects for our sample \citep[see also][]{ripepietal2018}.
This means that we dealt with cluster red giant stars placed above the GCs' horizontal branches.
To select cluster stars we constrained our sample to those satisfying the following criteria: i) stars 
located at the LMC distance,  i.e., $|\varpi|$ $<$ 3$\sigma(\varpi)$ 
\citep[see][]{vasiliev2018}; ii) 
stars located within the  tidal cluster radii taken from \citet{pm2018}. 

We then performed a maximum likelihood statistics \citep{pm1993,walker2006} in order to estimate
the mean  proper motions and dispersion for different subsets of stars, namely, those with 
$\sigma$(pmra)=$\sigma$(pmdec) $\le$ 0.1, 0.2, 0.3, 0.4 and 0.5 mas/yr, respectively. 
We optimised the probability $\mathcal{L}$ that a
given ensemble of stars with 
proper motions pm$_i$ and errors $\sigma_i$ are drawn from a population with mean
proper motion $<$pm$>$ and dispersion  W, 
as follows:\\

$\small
\mathcal{L}\,=\,\prod_{i=1}^N\,\left( \, 2\pi\,(\sigma_i^2 + W^2 \, ) 
\right)^{-\frac{1}{2}}\,\exp \left(-\frac{(pm_i \,- <pm>)^2}{2(\sigma_i^2 + W^2)}
\right) \\$

\noindent where the errors on the mean and dispersion were computed from the respective 
covariance matrices. We applied the above procerure for pmra and pmdec, separately.

Fig.~\ref{fig:fig2} depicts the results represented by ellipses centred on the mean values 
and with axes equal to the derived errors. As can be seen, the larger the individual
proper motion errors, the larger the derived errors of the mean proper motions, 
with some exception. With the aim of
assuring accuracy, we constrained the subsequent analysis to the GC mean proper motions  derived from 
stars with proper motion errors $\le$ 0.1 mas/yr (see Table~\ref{tab:table1}, where n refers to the number 
of stars used between those with proper motion errors 0.1 and 0.5 mas/yr, respectively). 
In order to illustrate the contamination of field stars in the GC
mean proper motion estimates, we have included with black thick dots every individual field stars
with proper motion errors $\le$ 0.1 mas/yr lying in the sky within a circle equivalent to the size of the cluster's circle, which is 
centred at 5 cluster radii away from the cluster centre. As far as we are aware, the 
resultant mean LMC GC proper motions represent those based on accurate measurements of 
{\it bonafide} cluster members.

\begin{figure*}
     \includegraphics[width=\columnwidth]{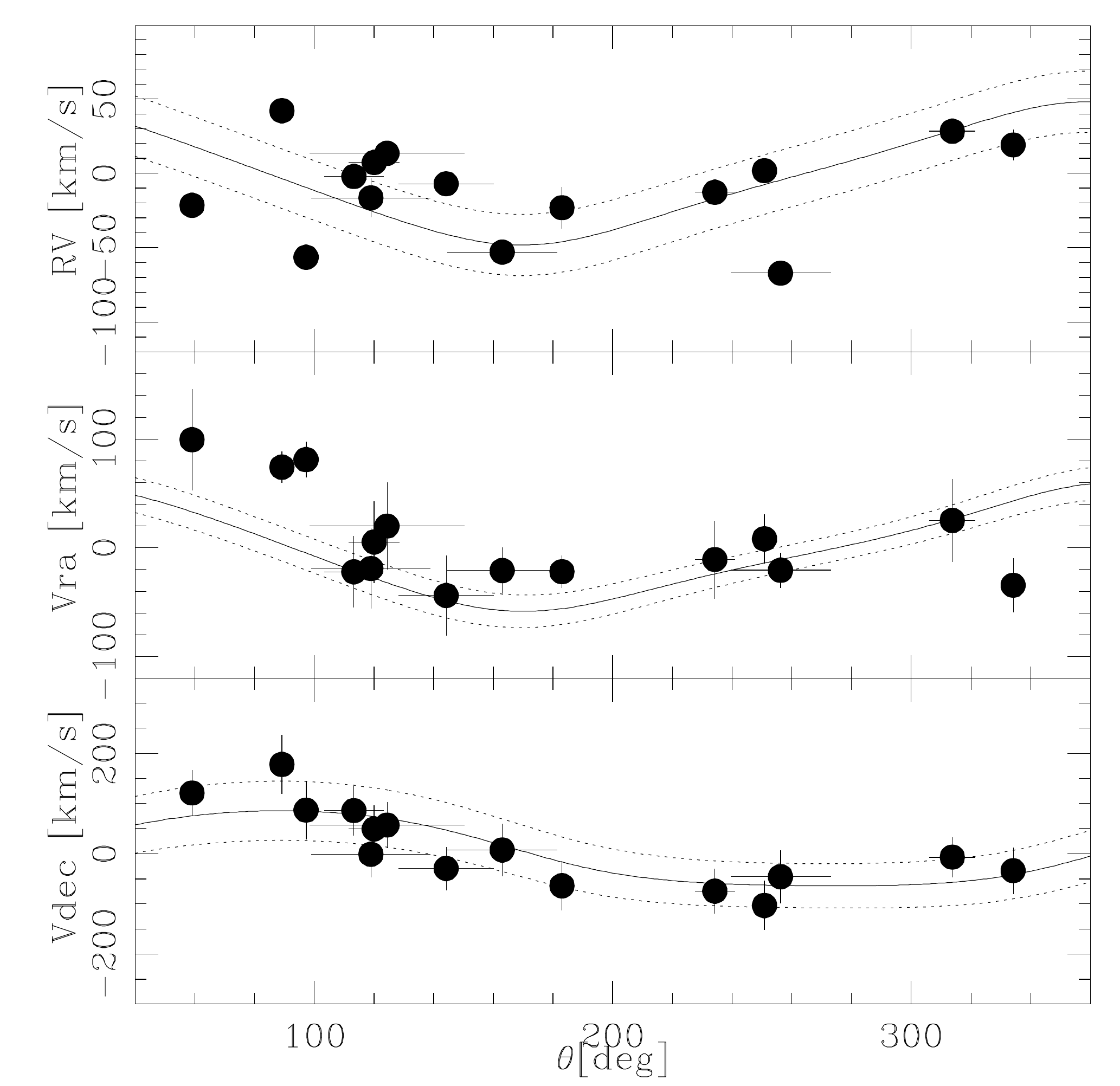}
     \includegraphics[width=\columnwidth]{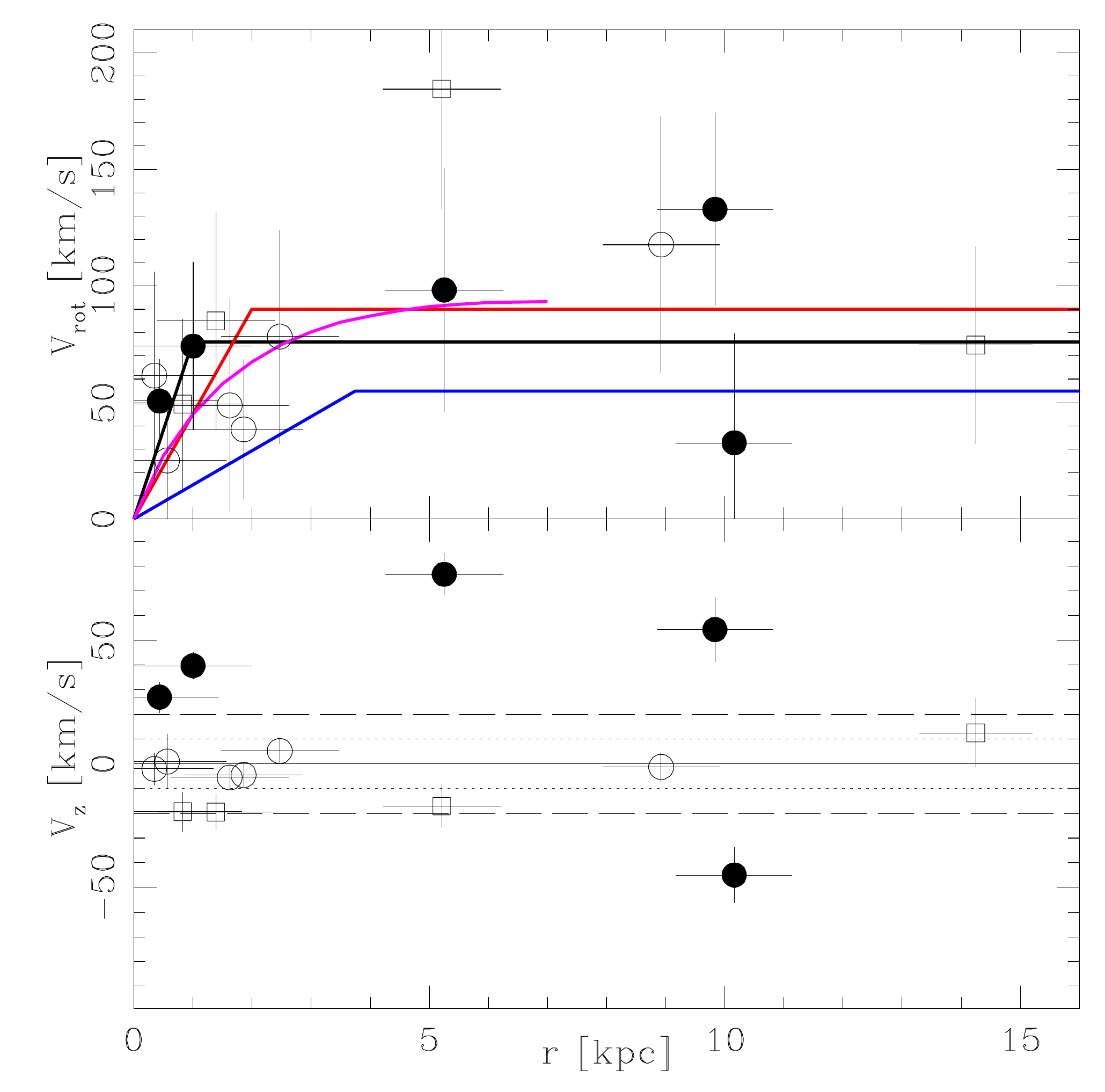}
     \includegraphics[width=\columnwidth]{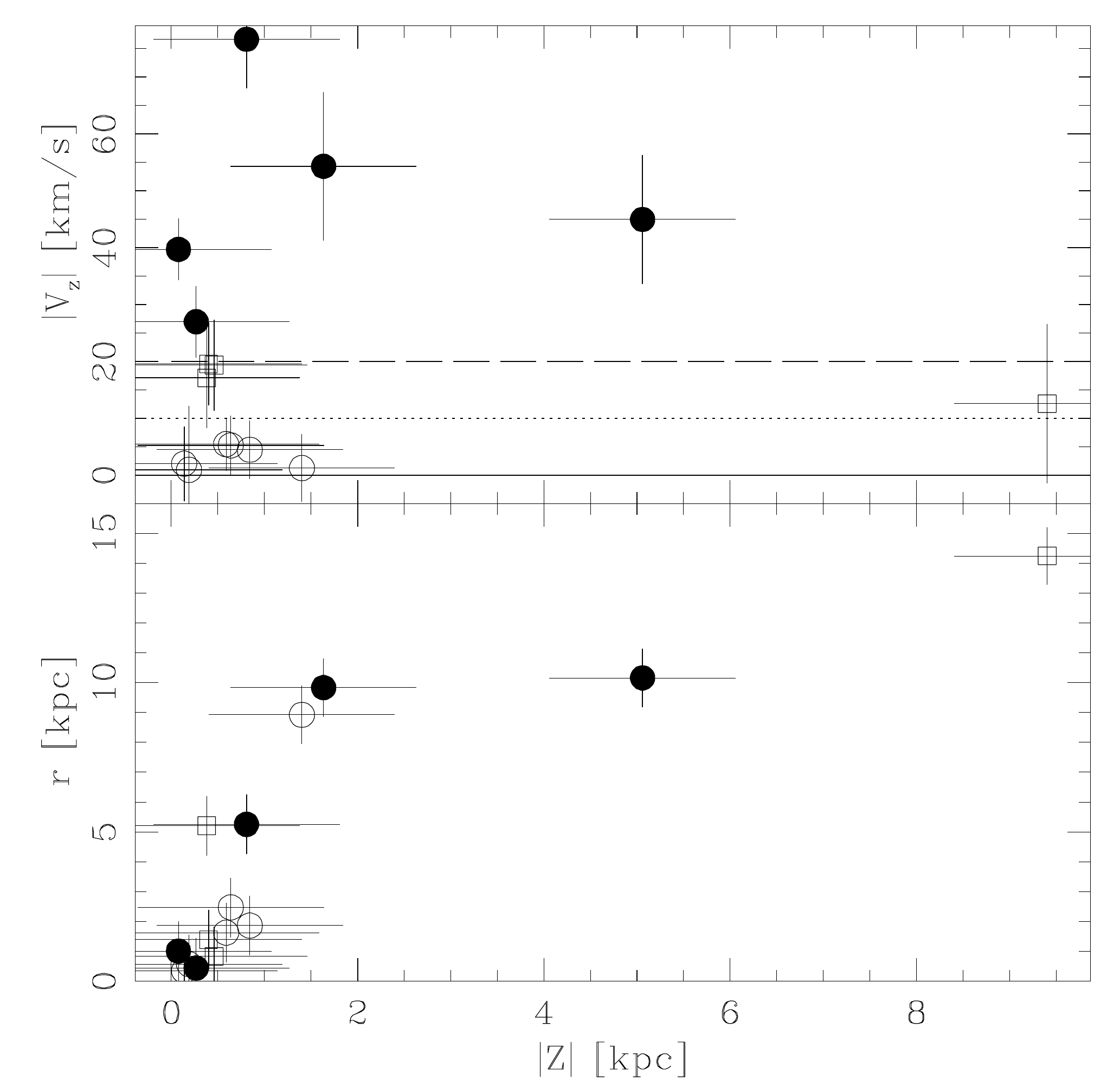}
     \includegraphics[width=\columnwidth]{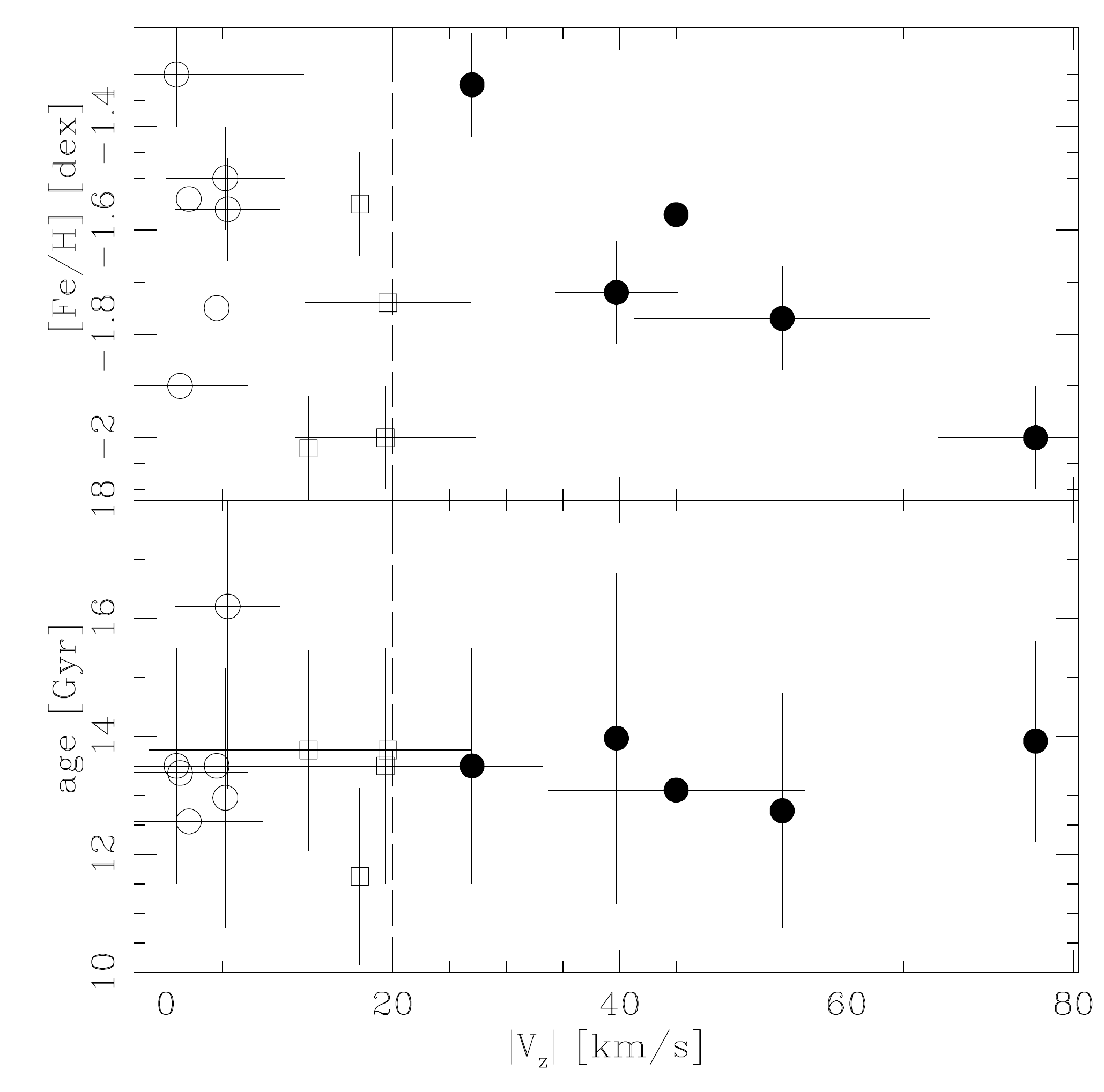}
   \caption{RVs and proper motions versus PAs diagrams of LMC GCs (top-left
panel). RVs and PAs were taken from \citet{piattietal2018}. We overplotted the
curve representing the best-fit solution for the LMC  disc rotation derived by \citet{vdmk14} from $HST$
proper motions of 22 LMC fields. The GC rotational and perpendicular velocities
as a function of their deprojected galactocentric distances ($r$) are shown
in the top-right panel. Open circles and  boxes represent GCs with vertical velocities 
$|$V$_z$$|$ $<$ 10 km/s  and 10 km/s $<$ $|$V$_z$$|$ $<$ 20 km/s, respectively.
The LMC rotation curves derived from $HST$ proper motions 
of 22 fields, and from line-of-sight velocities of young and old stellar 
populations are drawn with black, red and blue solid lines, respectively 
\citep[taken from Figure 7 of][]{vdmk14}. The LMC disc rotation curve
derived by \citet{vasiliev2018} is drawn with a magenta line.
Relationships of the height out of the plane ($|$Z$|$) with  $|$V$_z$$|$ 
and $r$ (bottom-left panel) and that of  $|$V$_z$$|$ with the GCs' ages and
metallicities (bottom-right panel) (see text for details). 
}
 \label{fig:fig1}
\end{figure*}

These proper motions can be compared with those of the LMC disc by adopting the transformation
equations (9), (13) and (21) in \citet{vdmareletal2002} and the best-fit solutions for the rotation
of the LMC disc obtained from $HST$ 
proper motions of field stars \citep[column 3 of Table 1 in][]{vdmk14}.  We then subtracted
the corresponding amount of motion of the LMC centre of mass from the GCs' proper motions. 
The top-left
panel of Fig.~\ref{fig:fig1} shows the results as a function of the PA. As a matter of units, we
used [km/s] = 4.7403885*$D_o$ [mas/yr], where $D_o$ is the distance to the LMC centre of mass 
\citep[= 50.1 kpc][]{vdmk14}, and denoted Vra and Vdec the movements in R.A and Dec., respectively. 
We have overplotted the rotation of the LMC disc with a solid line and 
those considering the errors in the inclination of the disc, the PA of the line-of-nodes, the systemic
and transversal velocities of the LMC centre of mass and disc velocity dispersion with dotted
lines, respectively. The errorbars of the plotted GCs' proper motions account not only for the 
measured errors listed in Table~\ref{tab:table1}, but also for those from the adopted 
\citet{vdmk14}'s best-fit solution for the 3D movement of the LMC centre of mass, propagated through the
transformation equations and added in quadrature. As can be seen, the GC motions relative to the LMC 
centre projected onto the sky resemble that of the rotation of a disc, with some noticeable scatter and some GCs placed beyond that rotational pattern.

\begin{table*}
\caption{Astrophysical properties of LMC GCs.}
\label{tab:table1}
\begin{tabular}{@{}lcccccccccc}\hline
ID          & pmra                & pmdec          & n  & $r$   &  $Z$  &  $V_{rot}$  & $V_z$   & class$^a$ & Age & [Fe/H] \\
            & (mas/yr)            & (mas/yr)       & & (kpc) &  (kpc) &  (km/s)     & (km/s)    &       & (Gyr) & (dex) \\\hline
NGC\,1466 & 1.769$\pm$0.074 & -0.571$\pm$0.050 & 8-105 & 8.92  &  1.40  &  117.8$\pm$55.2 & -1.2$\pm$5.9  & disc & 13.38$\pm$1.90 &-1.90$\pm$0.10 \\
NGC\,1754 & 1.947$\pm$0.048 & -0.177$\pm$0.063 & 19-86 & 2.47  &  0.63  &   78.3$\pm$45.9 &  5.2$\pm$5.2  & disc & 12.96$\pm$2.20 &-1.50$\pm$0.10  \\  
NGC\,1786 & 1.802$\pm$0.017 &  0.077$\pm$0.021 & 13-207& 1.86  & -0.84  &   38.5$\pm$29.9 & -4.4$\pm$5.1  & disc & 13.50$\pm$2.00 &-1.75$\pm$0.10  \\
NGC\,1835 & 1.994$\pm$0.013 & -0.005$\pm$0.024 & 23-341& 1.00  &  0.07  &   74.2$\pm$36.0 & 39.7$\pm$5.4  & halo & 13.97$\pm$2.80 &-1.72$\pm$0.10  \\
NGC\,1841 & 1.937$\pm$0.026 & -0.032$\pm$0.031 & 6-184 & 14.25 &  9.40  &   74.7$\pm$42.3 & 12.5$\pm$14.0 & disc & 13.77$\pm$1.70 &-2.02$\pm$0.10  \\ 
NGC\,1898 & 1.996$\pm$0.017 &  0.283$\pm$0.020 & 15-282& 0.43  &  0.26  &   50.6$\pm$17.9 & 26.9$\pm$6.2  & halo & 13.50$\pm$2.00 &-1.32$\pm$0.10  \\
NGC\,1916 & 1.828$\pm$0.055 &  0.494$\pm$0.083 & 27-165& 0.34  &  0.13  &   61.5$\pm$44.5 & -2.0$\pm$6.5  & disc & 12.56$\pm$5.50 &-1.54$\pm$0.10  \\
NGC\,1928 & 1.991$\pm$0.040 &  0.259$\pm$0.054 & 15-51 & 0.56  &  0.19  &   25.1$\pm$29.7 &  0.9$\pm$11.2 & disc & 13.50$\pm$2.00 &-1.30$\pm$0.10  \\  
NGC\,1939 & 2.092$\pm$0.023 &  0.163$\pm$0.028 & 6-197 & 0.83  &  0.45  &   49.3$\pm$36.3 &-19.3$\pm$7.9  & disc & 13.50$\pm$2.00 &-2.00$\pm$0.10  \\  
NGC\,2005 & 2.003$\pm$0.044 &  0.676$\pm$0.063 & 22-99 & 1.39  &  0.40  &   85.0$\pm$47.0 &-19.5$\pm$7.2  & disc & 13.77$\pm$4.90 &-1.74$\pm$0.10  \\
NGC\,2019 & 1.903$\pm$0.034 &  0.413$\pm$0.075 & 10-262& 1.62  &  0.58  &   48.8$\pm$45.8 & -5.4$\pm$4.6  & disc & 16.20$\pm$3.10 &-1.56$\pm$0.10  \\ 
NGC\,2210 & 1.578$\pm$0.030 &  1.212$\pm$0.025 & 7-139 & 5.21  &  0.38  &  184.6$\pm$51.7 &-17.0$\pm$8.8  & disc & 11.63$\pm$1.50 &-1.55$\pm$0.10  \\  
NGC\,2257 & 1.448$\pm$0.090 &  1.012$\pm$0.028 & 10-122& 9.83  & -1.63  &  132.9$\pm$41.3 & 54.3$\pm$13.0 & halo & 12.74$\pm$2.00 &-1.77$\pm$0.10  \\
Hodge\,11 & 1.541$\pm$0.023 &  0.861$\pm$0.028 & 7-200 & 5.25  &  0.80  &   98.2$\pm$52.4 & 76.6$\pm$8.6  & halo & 13.92$\pm$1.70 &-2.00$\pm$0.10  \\  
Reticulum & 1.984$\pm$0.035 & -0.350$\pm$0.035 & 7-98  & 10.16 & -5.05  &   32.7$\pm$47.0 &-44.9$\pm$11.3 & halo & 13.09$\pm$2.10 &-1.57$\pm$0.10  \\\hline 
\end{tabular}

\noindent $^a$ suggested GC population classification (see Section 3).
    
\end{table*}

\section{Analysis and discussion}

The projection of the GC velocity vectors onto the LMC plane and that perpendicular to it provide us
with a valuable tool to address the issue of the genuine rotation of them in the LMC disc 
and the orientation of their orbits around the LMC centre. In order to convert the vector (RV,Vra,Vdec)
into that with components $V_x$ and $V_y$ in the LMC plane and $V_z$ perpendicular to it, 
according to the reference system defined by \citet[][see their Figure 3]{vdmareletal2002}, we inverted
the matrix {\bf A = B $\times$ C}, where {\bf B} is the matrix :

\begin{eqnarray}
\left(
\begin{array}{ccc}
1 & 0 & 0 \\
0 & b_1  & b_2  \\
0 & b_3  & b_4   \\
\end{array}
\right)
\end{eqnarray}

\noindent with $b_1$, $b_2$, $b_3$ and $b_4$ being the coefficients of the transformation
equation (9) and {\bf C} the matrix defined in equation (5) of \citet{vdmareletal2002}, respectively, so that:

\begin{eqnarray}
\left(
\begin{array}{c}
V_x \\
V_y \\
V_z \\
\end{array}
\right) = 
{\bf A^{-1}}
\left(
\begin{array}{c}
{\rm RV} \\
{\rm Vra} \\
{\rm Vdec} \\
\end{array}
\right)
\end{eqnarray}

\noindent Hence, $V_{rot}$ = ($V_x^2$ + $V_y^2$)$^{1/2}$, while the errors $\sigma$($V_x$), $\sigma$($V_y$)
and $\sigma$($V_z$) were computing from propagation of errors of eq. (2). The resulting values for $V_{rot}$
and $V_z$ are listed in Table~\ref{tab:table1}.

The top-right panel of Fig.~{\ref{fig:fig1} shows the resulting relationships of $V_{rot}$ and $V_z$
as a function of the deprojected distances ($r$). At first glance, inner GCs ($r <$ 5 kpc) seem to have
$V_{rot}$ values in  better  agreement with the rotation curve of the LMC disc than those in the outer regions.
However, by inspecting the $V_z$ versus $r$ diagram, it is possible to distinguish two groups of GCs: one
group with an average velocity dispersion perpendicular to the LMC plane close to zero, and another group 
with $|V_z| >$ 10 km/s (another plausible cut could be at 20 km/s). For the sake of the reader, we have drawn GCs with $|V_z|$ smaller than 10 km/s, between 10 and 20 km/s and larger  than
20 km/s with open circles and boxes and filled circles, respectively, and traced  those limits with  dotted and dashed lines. Notice that, The LMC rotation curve derived from old field stars (blue line) by \citet{vdmk14} does not
particularly agree with GCs. Other curves (e.g. black, red, magenta lines) seem to agree better with GCs across the range of $r$. GCs with
a relatively high $V_z$ velocity are those that globally more depart from the LMC rotation curve.
For instance, the mean
difference (absolute value) between the GC $V_{rot}$ values and those for the same $r$ values along the mean LMC disc rotation curve
(magenta line) turned out to be 13.4$\pm$16.1,  24.8$\pm$21.1 and 28.3$\pm$13.0 for 
GCs with $|V_z|$ $<$ 10 km/s, 10 km/s $<$ $|$V$_z$$|$ $<$ 20 km/s and $|V_z|$ $>$ 20 km/s
(31.7$\pm$12.1 for $|V_z|$ $>$ 10 km/s and 18.8$\pm$12.6 for $|V_z|$ $<$ 20 km/s), respectively. If we consider only GCs with $r$ $<$ 5 kpc, we get 14.5$\pm$16.2, 15.5$\pm$28.6 and 30.0$\pm$15.3 (26.7$\pm$13.5 for $|V_z|$ $>$ 10 km/s and
14.7$\pm$14.2 for $|V_z|$ $<$ 20 km/s), respectively.
 
Such a kinematic distinction has also its counterpart in the spatial distribution of the GCs. The
bottom-left panel shows that the farther a GC from the LMC centre, the larger its height out of the
plane. Along this trend, the GCs with $|V_z| <$ 10 km/s are those confined to the LMC disc, with the
sole exception of NGC\,1466 ($r$=8.92 kpc). This appears to be a peculiar GC, because it has a
relatively large $V_{rot}$, even though its $|V_z|$ is smaller than 10 km/s. 
If we considered GCs with $|V_z| <$ 20 km/s, the trend would remain with the additional exception of NGC\,1841  ($r$=9.40 kpc).
GCs with  $|V_z|$ $<$ 10km/s expand the $r$ range $\sim$ 0 - 3 kpc (0 - 9 kpc if NGC\,1466 is included), those with $|V_z|$ $<$ 20 km/s expand the $r$ range $\sim$ 0 - 5 kpc (0 - 15 kpc if NGC\,1481 is included), while
those with $|V_z|$ $>$ 20 km/s reached $r$ $\sim$ 10 kpc.
As the height out of the LMC plane is
considered, GCs with $|V_z|$ values smaller and larger than 10 km/s are located at $|Z|$  values
smaller than 0.5 kpc (1.5 kpc if NGC\,1466 is included) and 8.6 kpc, respectively. 
Z values were calculated from eq. (7) in \citet{vdmc2001}, with distances taken from \citet{wagnerkaiseretal2017} and \citet{pm2018}, and assuming distance errors of 2 kpc, i.e. twice 
as big the dispersion of the GCs' distances obtained by \citet{wagnerkaiseretal2017}.
Both spatial regimes tell us also about two different spatial patterns, one closely related to a
disc component and another more similar to an spherical one. \citet{bekki2007} performed numerical
simulations of the LMC formation with the aim of looking for an answer to the, until then, kinematic
difference between LMC halo field stars and GCs. Surprisingly, he found that GCs have little
rotation and spatial distribution and kinematics similar to those of the halo stars. Our analysis,
based on the  velocity vectors of GCs with high $V_z$ agree very well with that theoretical result.

Finally, we analysed whether there is any link of these  two phase-space GC populations  with their ages and metallicities, 
so that some clues about their origins can be inferred. The analysis includes the two $|V_z|$ cuts.
The bottom-right panel of Fig.~\ref{fig:fig1}
shows the resultant behaviours, where ages and metallicities were taken from \citet{pm2018} and
\citet{piattietal2018}  (see Table~\ref{tab:table1}). From a maximum likelihood analysis we found that  these groups 
of GCs are coeval at a level of 0.4 Gyr and that they hardly differ in metallicity.
This  means that both GC components
have formed nearly at the same time, within $\sim$ 3 Gyr of their formation 
\citep[12 $\la$ age (Gyr) $\la$ 14,][]{piattietal2009,wagnerkaiseretal2018} and under similar
chemical enrichment processes (the three groups expand similar [Fe/H] ranges). Because of these
similarities in age and metallicity and the noticeable difference in their spatial
distributions (thin disc and extended halo), we speculate with the possibility that their
origins could have occurred through a fast collapse that formed halo and disc concurrently.
The inner GCs with $|V_z|<$ 10 km/s have remained rotating in the LMC disc since their
{\it in-situ} formation. As for the formation of the GCs with higher $V_z$ values, they could have
occurred {\it in-situ} in the LMC halo and/or stripped from the Small Magellanic Cloud (SMC), 
where GCs with similar ages and
metallicities should have formed. As suggested by \citet{carpinteroetal2013}, this could be a
plausible explanation for the lack of old metal-poor GCs in the SMC. As for the time of the GC stripping,
we do not have any hint: it could have happened at the early formation of both galaxies or during a
later approach of the SMC by the LMC.

\begin{figure*}
     \includegraphics[width=\textwidth]{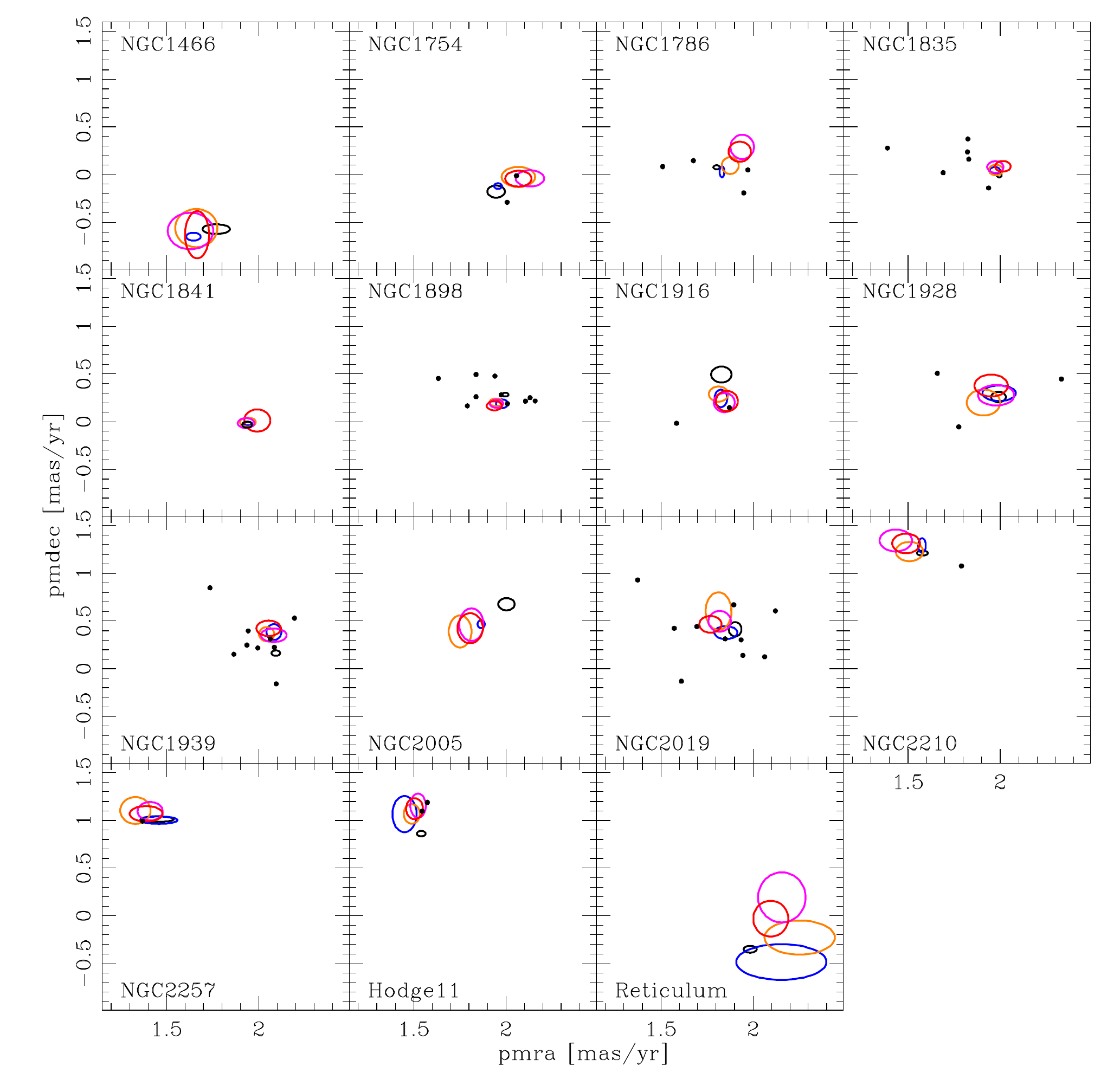}
   \caption{Vector-point diagrams of LMC GCs' mean proper motions
derived using {\it Gaia} DR2 data with $\sigma$(pmra)=$\sigma$(pmdec) $\le$ 
0.1, 0.2, 0.3, 0.4 and 0.5 mas/yr, represented by black, blue, orange, magenta and
red ellipses, respectively. Black points represent individual field stars located
in an equal circular cluster area centred at 5 cluster radius from the cluster centre
with proper motions errors $\le$ 0.1 mas/yr (see text for details).}
 \label{fig:fig2}
\end{figure*}

\section*{Acknowledgements}
We thank the referee for the thorough reading of the manuscript and
timely suggestions to improve it. We thanks Eugene Vasiliev his valuable comments 
on a draft version of this work. EJA acknowledges financial support from 
MINECO (Spain) through grant AYA2016-75931-C2-1-P.
This work has made use of data from the European Space Agency (ESA) mission
{\it Gaia} (\url{https://www.cosmos.esa.int/gaia}), processed by the {\it Gaia}
Data Processing and Analysis Consortium (DPAC,
\url{https://www.cosmos.esa.int/web/gaia/dpac/consortium}). Funding for the DPAC
has been provided by national institutions, in particular the institutions
participating in the {\it Gaia} Multilateral Agreement.




\bibliographystyle{mnras}

\begin{thebibliography}{}
\makeatletter
\relax
\def\mn@urlcharsother{\let\do\@makeother \do\$\do\&\do\#\do\^\do\_\do\%\do\~}
\def\mn@doi{\begingroup\mn@urlcharsother \@ifnextchar [ {\mn@doi@}
  {\mn@doi@[]}}
\def\mn@doi@[#1]#2{\def\@tempa{#1}\ifx\@tempa\@empty \href
  {http://dx.doi.org/#2} {doi:#2}\else \href {http://dx.doi.org/#2} {#1}\fi
  \endgroup}
\def\mn@eprint#1#2{\mn@eprint@#1:#2::\@nil}
\def\mn@eprint@arXiv#1{\href {http://arxiv.org/abs/#1} {{\tt arXiv:#1}}}
\def\mn@eprint@dblp#1{\href {http://dblp.uni-trier.de/rec/bibtex/#1.xml}
  {dblp:#1}}
\def\mn@eprint@#1:#2:#3:#4\@nil{\def\@tempa {#1}\def\@tempb {#2}\def\@tempc
  {#3}\ifx \@tempc \@empty \let \@tempc \@tempb \let \@tempb \@tempa \fi \ifx
  \@tempb \@empty \def\@tempb {arXiv}\fi \@ifundefined
  {mn@eprint@\@tempb}{\@tempb:\@tempc}{\expandafter \expandafter \csname
  mn@eprint@\@tempb\endcsname \expandafter{\@tempc}}}

\bibitem[\protect\citeauthoryear{{Bekki}}{{Bekki}}{2007}]{bekki2007}
{Bekki} K.,  2007, \mn@doi [\mnras] {10.1111/j.1365-2966.2007.12219.x}, \href
  {http://adsabs.harvard.edu/abs/2007MNRAS.380.1669B} {380, 1669}

\bibitem[\protect\citeauthoryear{{Bekki}}{{Bekki}}{2011}]{bekki2011}
{Bekki} K.,  2011, \mn@doi [\mnras] {10.1111/j.1365-2966.2011.19211.x}, \href
  {http://adsabs.harvard.edu/abs/2011MNRAS.416.2359B} {416, 2359}

\bibitem[\protect\citeauthoryear{{Brocato}, {Castellani}, {Ferraro},
  {Piersimoni}  \& {Testa}}{{Brocato} et~al.}{1996}]{brocatoetal1996}
{Brocato} E.,  {Castellani} V.,  {Ferraro} F.~R.,  {Piersimoni} A.~M.,
  {Testa} V.,  1996, \mn@doi [\mnras] {10.1093/mnras/282.2.614}, \href
  {http://adsabs.harvard.edu/abs/1996MNRAS.282..614B} {282, 614}

\bibitem[\protect\citeauthoryear{{Carpintero}, {G{\'o}mez}  \&
  {Piatti}}{{Carpintero} et~al.}{2013}]{carpinteroetal2013}
{Carpintero} D.~D.,  {G{\'o}mez} F.~A.,   {Piatti} A.~E.,  2013, \mn@doi
  [\mnras] {10.1093/mnrasl/slt096}, \href
  {http://adsabs.harvard.edu/abs/2013MNRAS.435L..63C} {435, L63}

\bibitem[\protect\citeauthoryear{{Gaia Collaboration} et~al.,}{{Gaia
  Collaboration} et~al.}{2016}]{gaiaetal2016}
{Gaia Collaboration} et~al., 2016, \mn@doi [\aap]
  {10.1051/0004-6361/201629272}, \href
  {http://adsabs.harvard.edu/abs/2016A%26A...595A...1G} {595, A1}

\bibitem[\protect\citeauthoryear{{Gaia Collaboration} et~al.,}{{Gaia
  Collaboration} et~al.}{2018a}]{gaiaetal2018b}
{Gaia Collaboration} et~al., 2018a, \mn@doi [\aap]
  {10.1051/0004-6361/201833051}, \href
  {http://adsabs.harvard.edu/abs/2018A%26A...616A...1G} {616, A1}

\bibitem[\protect\citeauthoryear{{Gaia Collaboration} et~al.,}{{Gaia
  Collaboration} et~al.}{2018b}]{gaiaetal2018a}
{Gaia Collaboration} et~al., 2018b, \mn@doi [\aap]
  {10.1051/0004-6361/201832698}, \href
  {http://adsabs.harvard.edu/abs/2018A%26A...616A..12G} {616, A12}

\bibitem[\protect\citeauthoryear{{Grocholski}, {Cole}, {Sarajedini}, {Geisler}
  \& {Smith}}{{Grocholski} et~al.}{2006}]{getal06}
{Grocholski} A.~J.,  {Cole} A.~A.,  {Sarajedini} A.,  {Geisler} D.,   {Smith}
  V.~V.,  2006, \mn@doi [\aj] {10.1086/507303}, 132, 1630

\bibitem[\protect\citeauthoryear{{Lindegren} et~al.,}{{Lindegren}
  et~al.}{2018}]{lindegrenetal2018}
{Lindegren} L.,  et~al., 2018, \mn@doi [\aap] {10.1051/0004-6361/201832727},
  \href {http://adsabs.harvard.edu/abs/2018A%26A...616A...2L} {616, A2}

\bibitem[\protect\citeauthoryear{{Piatti} \& {Geisler}}{{Piatti} \&
  {Geisler}}{2013}]{pg13}
{Piatti} A.~E.,  {Geisler} D.,  2013, \mn@doi [\aj]
  {10.1088/0004-6256/145/1/17}, 145, 17

\bibitem[\protect\citeauthoryear{{Piatti} \& {Mackey}}{{Piatti} \&
  {Mackey}}{2018}]{pm2018}
{Piatti} A.~E.,  {Mackey} A.~D.,  2018, \mn@doi [\mnras]
  {10.1093/mnras/sty1048}, \href
  {http://adsabs.harvard.edu/abs/2018MNRAS.tmp..991P} {}

\bibitem[\protect\citeauthoryear{{Piatti}, {Geisler}, {Sarajedini}  \&
  {Gallart}}{{Piatti} et~al.}{2009}]{piattietal2009}
{Piatti} A.~E.,  {Geisler} D.,  {Sarajedini} A.,   {Gallart} C.,  2009, \mn@doi
  [\aap] {10.1051/0004-6361/200912223}, \href
  {http://adsabs.harvard.edu/abs/2009A%26A...501..585P} {501, 585}

\bibitem[\protect\citeauthoryear{{Piatti}, {Hwang}, {Cole}, {Angelo}  \&
  {Emptage}}{{Piatti} et~al.}{2018}]{piattietal2018}
{Piatti} A.~E.,  {Hwang} N.,  {Cole} A.~A.,  {Angelo} M.~S.,   {Emptage} B.,
  2018, \mn@doi [\mnras] {10.1093/mnras/sty2324}, \href
  {http://adsabs.harvard.edu/abs/2018MNRAS.481...49P} {481, 49}

\bibitem[\protect\citeauthoryear{{Pryor} \& {Meylan}}{{Pryor} \&
  {Meylan}}{1993}]{pm1993}
{Pryor} C.,  {Meylan} G.,  1993, in {Djorgovski} S.~G.,  {Meylan} G.,  eds,
  Astronomical Society of the Pacific Conference Series Vol. 50, Structure and
  Dynamics of Globular Clusters. p.~357

\bibitem[\protect\citeauthoryear{{Ripepi}, {Molinaro}, {Musella}, {Marconi},
  {Leccia}  \& {Eyer}}{{Ripepi} et~al.}{2018}]{ripepietal2018}
{Ripepi} V.,  {Molinaro} R.,  {Musella} I.,  {Marconi} M.,  {Leccia} S.,
  {Eyer} L.,  2018, preprint, \href
  {http://adsabs.harvard.edu/abs/2018arXiv181010486R} {} (\mn@eprint {arXiv}
  {1810.10486})

\bibitem[\protect\citeauthoryear{{Schommer}, {Suntzeff}, {Olszewski}  \&
  {Harris}}{{Schommer} et~al.}{1992}]{s92}
{Schommer} R.~A.,  {Suntzeff} N.~B.,  {Olszewski} E.~W.,   {Harris} H.~C.,
  1992, \mn@doi [\aj] {10.1086/116074}, \href
  {http://adsabs.harvard.edu/abs/1992AJ....103..447S} {103, 447}

\bibitem[\protect\citeauthoryear{{Sharma}, {Borissova}, {Kurtev}, {Ivanov}  \&
  {Geisler}}{{Sharma} et~al.}{2010}]{shetal10}
{Sharma} S.,  {Borissova} J.,  {Kurtev} R.,  {Ivanov} V.~D.,   {Geisler} D.,
  2010, \mn@doi [\aj] {10.1088/0004-6256/139/3/878}, 139, 878

\bibitem[\protect\citeauthoryear{{Vasiliev}}{{Vasiliev}}{2018}]{vasiliev2018}
{Vasiliev} E.,  2018, \mn@doi [\mnras] {10.1093/mnrasl/sly168}, \href
  {http://adsabs.harvard.edu/abs/2018MNRAS.481L.100V} {481, L100}

\bibitem[\protect\citeauthoryear{{Wagner-Kaiser} et~al.,}{{Wagner-Kaiser}
  et~al.}{2017}]{wagnerkaiseretal2017}
{Wagner-Kaiser} R.,  et~al., 2017, \mn@doi [\mnras] {10.1093/mnras/stx1702},
  \href {http://adsabs.harvard.edu/abs/2017MNRAS.471.3347W} {471, 3347}

\bibitem[\protect\citeauthoryear{{Wagner-Kaiser}, {Mackey}, {Sarajedini},
  {Cohen}, {Geisler}, {Yang}, {Grocholski}  \& {Cummings}}{{Wagner-Kaiser}
  et~al.}{2018}]{wagnerkaiseretal2018}
{Wagner-Kaiser} R.,  {Mackey} D.,  {Sarajedini} A.,  {Cohen} R.~E.,  {Geisler}
  D.,  {Yang} S.-C.,  {Grocholski} A.~J.,   {Cummings} J.~D.,  2018, \mn@doi
  [\mnras] {10.1093/mnras/stx3061}, \href
  {http://adsabs.harvard.edu/abs/2018MNRAS.474.4358W} {474, 4358}

\bibitem[\protect\citeauthoryear{{Walker}, {Mateo}, {Olszewski}, {Bernstein},
  {Wang}  \& {Woodroofe}}{{Walker} et~al.}{2006}]{walker2006}
{Walker} M.~G.,  {Mateo} M.,  {Olszewski} E.~W.,  {Bernstein} R.,  {Wang} X.,
  {Woodroofe} M.,  2006, \mn@doi [\aj] {10.1086/500193}, \href
  {http://adsabs.harvard.edu/abs/2006AJ....131.2114W} {131, 2114}

\bibitem[\protect\citeauthoryear{{van der Marel} \& {Cioni}}{{van der Marel} \&
  {Cioni}}{2001}]{vdmc2001}
{van der Marel} R.~P.,  {Cioni} M.-R.~L.,  2001, \mn@doi [\aj]
  {10.1086/323099}, \href {http://adsabs.harvard.edu/abs/2001AJ....122.1807V}
  {122, 1807}

\bibitem[\protect\citeauthoryear{{van der Marel} \& {Kallivayalil}}{{van der
  Marel} \& {Kallivayalil}}{2014}]{vdmk14}
{van der Marel} R.~P.,  {Kallivayalil} N.,  2014, \mn@doi [\apj]
  {10.1088/0004-637X/781/2/121}, 781, 121

\bibitem[\protect\citeauthoryear{{van der Marel}, {Alves}, {Hardy}  \&
  {Suntzeff}}{{van der Marel} et~al.}{2002}]{vdmareletal2002}
{van der Marel} R.~P.,  {Alves} D.~R.,  {Hardy} E.,   {Suntzeff} N.~B.,  2002,
  \mn@doi [\aj] {10.1086/343775}, \href
  {http://adsabs.harvard.edu/abs/2002AJ....124.2639V} {124, 2639}

\makeatother
\end{thebibliography}

\input{paper.bbl}







\bsp	
\label{lastpage}
\end{document}